\documentclass[prb,showpacs,twocolumn]{revtex4}
\usepackage{graphicx,amssymb,amsmath,psfrag,color}
\begin{document}
\title{Local Classical and Quantum Criticality due to Electron-Vibration Interaction}
\author{Bal\'azs D\'ora}
\email{dora@pks.mpg.de}
\affiliation{Max-Planck-Institut f\"ur Physik komplexer Systeme, N\"othnitzer Str. 38, 01187 Dresden, Germany}

\date{\today}

\begin{abstract}
We study the local classical and quantum critical properties of electron-vibration interaction, represented by the Yu-Anderson model. 
It exhibits an instability, similar to the Wentzel-Bardeen singularity, whose
 nature resembles to weakly first order quantum phase transitions at low temperatures, and crosses over to Gaussian behaviour 
with increasing temperature. We determine the dominant energy scale separating the quantum from classical 
criticality, study the effect of dissipation and analyze its impact on correlation functions. 
Similar phenomenon should be observable in carbon nanotubes around local defects.

\end{abstract}


\maketitle

\section{Introduction}

Quantum phase transitions are intensively studied due to the governing fundamental physical properties and also because a number of
highly interesting systems display such behaviour. 
These occur at zero temperature at quantum critical points (QCP), and are dominated by quantum rather than
classical fluctuations. Their study enriches our knowledge on classical and 
quantum critical behaviour, and can
reveal the connections between them in terms of quantum-to-classical mappings\cite{sachdev}. 
Quantum phase transitions were found to explain the behaviour of Ge doped YbRh$_2$Si$_2$\cite{custers} 
and other heavy fermion materials\cite{qsi}.  

The notion "local criticality" stands for quantum phase transitions in e.g. quantum impurity models which evolve only in time 
but are confined in space, such as the 
quantum phase transition in the sub-Ohmic and Ohmic spin-boson model\cite{rmpleggett,weiss}, the dissipation induced phase transition 
in a quantum box\cite{lehur} or the local quantum phase transition in the pseudogap Anderson model\cite{glossop}.

Local vibrational modes due to foreign (adsorbed) molecules or lattice imperfections strongly influence the electronic transport and 
induce 
dephasing through inelastic scattering\cite{doraelastic}.  
Molecular electronic devices\cite{park,LeRoy} are probed and controlled locally by single-molecule vibrational spectroscopy\cite{stipe} 
based on STM 
and inelastic electron 
tunneling 
spectroscopy, and the spectrum of molecular vibrations often indicates more 
complex behaviour than can be seen in bulk transport\cite{galperinscience}, such as negative differential resistance and 
hysteresis\cite{lamagna}.
Conductance measurements on mechanically controllable break junctions reveal the presence of local vibrational degrees of freedom, 
when noble metal (Pt) electrodes were connected by a single molecule as Pt or H$_2$\cite{djukic,csonka}.
The vibrational mode softens after coupling it to electrons, 
and the vibrational resonance in the conductance due these modes shifts to lower energies. 
A critical bosonic mode caused by the interplay of softening and dissipation contributes to transport down to very low 
temperatures.
Thus, the detailed understanding of these modes and their local criticality beyond the mean field is 
essential\cite{galperinscience}, since even away from the critical point, they leave their mark on the responses\cite{sachdev}.

Recently the observation of strong phonon modes were reported in suspended carbon nanotubes (CN) by STM\cite{LeRoy}.
The electronic properties of CN are tunable by chemical doping or by changing the chirality of the 
tube, hence these systems are ideal candidates to study and control enhanced molecular vibrations\cite{LeRoy,park}, developing around 
lattice imperfections or encapsulated molecules via a local deformation potential.
Criticality due to local electron-vibration interaction is regarded as the descendant of the Wentzel-Bardeen (WB) 
singularity\cite{wentzel,bardeen,kohn}, which arises in a one dimensional system of electrons, 
coupled to long wavelength phonons, which only allows for forward scattering. 
For a critical value of the electron-phonon coupling, the system becomes
unstable and acquires a negative compressibility.
The thermodynamic
quantities  and correlation functions for the electron-phonon system
were studied near this singular point \cite{varga,loss}, indicating
the presence of a phase transition: the divergence of the specific heat is
accompanied by a collapse of the system induced by the strong
electron--phonon interaction.
For a critical value of the electron-phonon coupling, the system becomes
unstable and acquires a negative compressibility. A recent study\cite{eggerwb} suggested that the WB singularity could be reached 
experimentally in thick carbon nanotubes due to phonons.

Local vibrational modes and their critical properties are interesting for a variety of other reasons: 
critical modes are important in spintronics since they unavoidably lead to decoherence even at very low temperatures.
Electron-vibration interactions are exceptionally important for molecular solids where highly energetic vibrational states greatly
influence the electronic properties. Examples include the superconductivity with $T_c=40$~K in fullerides\cite{GunnRMP1997}
and the energetic electron-phonon sidebands in the excitonic excitation states of single-wall carbon nanotubes\cite{plentz}.

This paper addresses the local criticality caused by electron-vibration interaction. We determine the critical 
exponents, 
and show that a weakly first order quantum phase transition governs the low temperature physics, similarly to certain Ising models. 
At high temperatures, it crosses over to classical, Gaussian behaviour. This crossover influences the electronic properties as 
well e.g. in carbon nanotubes, which can be revealed by local spectroscopical measurements\cite{stipe}.

\section{The model and its basic properties}

As schematization of local electron-vibration interaction, we start with the Yu-Anderson or single impurity Holstein 
model\cite{yuanderson}:
\begin{equation}
H=\sum_{k}\varepsilon(k)c^+_{k}c_{k}+g_{d}Q\Psi^+({\bf 0})\Psi({\bf 0})+\frac{P^2}{2m}+\frac
{m\omega_0^2}{2}Q^2,
\label{hamilton}
\end{equation}
which describes d-dimensional electrons interacting with a local bosonic mode at a single impurity site with position $Q$ and 
momentum $P$. The model can be 
mapped onto one dimensional chiral 
fermions interacting with a single vibrational mode, and the fermionic field can be bosonized\cite{doraphonon,doraelastic}. Then, we 
arrive to an effective model of one dimensional coupled harmonic oscillators, i.e. the 
Caldeira-Leggett (CL) model\cite{caldeira,weiss}:
\begin{equation}
H=v_c\int\limits_{-\infty}^{\infty}\textmd{d}x
\left[\partial_x\Phi(x)\right]^2+\frac{g}{\sqrt\pi}Q\partial_x\Phi(0)+\frac{P^2}{2m}+\frac{m\omega_0^2}{2}Q^2,
\label{hamboson}
\end{equation}
$v_c$ is the charge velocity, and $g$ is the phase shift caused by $g_d$, $\Phi(x)$ stems from the bosonic representation of the 
fermion field. 
This also represents the effective model for the large spin-boson model\cite{brandes}.
After integrating out the bosonized electron field, $\Phi(x)$, the effective action for the phonon reads as
\begin{equation}
S_{ph}=\frac {m}{2T}\sum_n\left(\omega_n^2+\omega_0^2\left(1-\frac{\Gamma}{\Gamma_2}\right)+2|\omega_n|\Gamma\right)|Q_n|^2,
\label{effaction}
\end{equation}
where $\omega_n=2\pi nT$ is  the bosonic Matsubara frequency, $Q_n$'s are the Fourier components of $Q(\tau)$.
The main difference with respect to CL is the potential renormalization (the 
$-\omega_0^2\Gamma/\Gamma_2$ term), which is avoided in 
CL to study the effect of pure 
dissipation. In our case, the phonon is expected to soften after  coupling it to electrons
 on physical ground, therefore such local 
term is present in the action\cite{weiss}.
As the phonon mode softens, its eigenfrequencies on the real frequency axis are given by \cite{doraphonon}
\begin{equation}
\omega_{p\pm}=-i\Gamma\pm\sqrt{\omega_0^2(1-\Gamma/\Gamma_2)-\Gamma^2},
\end{equation}
where $\Gamma_2=\pi\omega_0^2/4W\ll\omega_0\ll W$, $W$ is the bandwidth of the conduction electrons, and
$\Gamma=\pi(g\rho)^2/2m$ for small $g$, and approaches $\Gamma_2$ as $g\rightarrow\infty$.
Here, $\rho=1/2\pi v_c$ is the chiral electron density of states.
The explicit dependence of $\Gamma$ on $g_d$ cannot be determined by the
bosonization approach\cite{doraphonon}. The real part of the phonon frequency remains finite (underdamped) for
$\Gamma<\Gamma_1\approx \Gamma_2(1-\Gamma_2^2/\omega_0^2)$. For
$\Gamma_1<\Gamma<\Gamma_2$,
the oscillatory behaviour disappears from the phononic response (Re$\omega_{p\pm}=0$), and two distinct
dampings characterize it (overdamped).

\section{Quantum criticality}

Close to $\Gamma_2$, the softening of the phonon frequency occurs as
\begin{equation}
\omega_{p+}= -\frac{i\omega_0^2}{2\Gamma_2}y,
\label{domenergy}
\end{equation}
where $y>0$ is the distance from criticality (the effective reduced "temperature"):
\begin{equation}
y=1-\frac{\Gamma}{\Gamma_2},
\end{equation}
and following Ref. \onlinecite{sachdev}, the characteristic energy scale is expected to vanish as $\omega_{p+}\sim y^\nu$.
This defines $\nu=1$, and using\cite{cardy} $\nu=1/y_t$, 
the "thermal" scaling exponent is $y_t=1$.
The extra $i$ in Eq. \eqref{domenergy} signals the dissipative nature of the transition.
Note, that criticality is tuned by $\Gamma$ and not by the temperature, so $y_t$ belongs to $\Gamma$.
At the same time, $\omega_{p-}$ approaches $-2i\Gamma_2$.

Since the problem is effectively one dimensional, possesses "0" spatial dimension and evolves only in time ($z=1$), this allows us 
to set $d_{eff}=d+z=1$. 
The dynamical 
exponent, $z$ can be melted in the definition of $d$, since there are no separate spatial and temporal dimensions. 

To proceed with the exploration of the critical properties of our model,
we evaluate its free energy. Using Eq. \eqref{effaction} or following Ref. \onlinecite{doraphonon}, it is obtained as
\begin{gather}
F=F_e+2\Gamma+\nonumber\\
+\frac {T}{\pi}\int \textmd{d}x \ln\left|1-\exp\left(-\frac x 
T\right)\right|\frac{\Gamma(x^2+\omega_0^2y)}{(x^2-\omega_0^2y)^2+(2\Gamma x)^2},
\label{freeenQ}
\end{gather}
where $F_e$ is the phonon free contribution of electrons.
Focusing on the most singular contribution in $y$ at $T=0$, we get
\begin{equation}
f_s(T\rightarrow 0)=\frac{\omega_0^2}{4\pi\Gamma_2} y \ln\left(\frac 1y\right)\sim y^{2-\alpha},
\end{equation}
leading to $2-\alpha=1=d_{eff}/y_t$ (using $\ln(x)=\lim_{\varepsilon\rightarrow 0} (x^\varepsilon-1)/\varepsilon$), from which 
$y_t=1$ is deduced in accordance 
with the exponent 
we got from the vanishing of the characteristic energy scale 
in Eq. \eqref{domenergy}. 
This suggests the emergence of a weakly first order quantum phase transition\cite{continento}.
Interestingly, many experimentally observed quantum phase transitions
 belong to this 
category\cite{pfleiderer}.
In contrast to true first order transitions, where $f_s\sim |y|$, we have 
logarithmic corrections in $y$, causing the second 
derivative of $f_s$ (the equivalent of the specific heat) to diverge in a power-law fashion as $1/y$.
True first order transitions are not accompanied by critical fluctuations. However, the weakly first order nature of the transition 
here 
allows for criticality to develop, which leaves its mark on the fluctuations. The mean square of the $Q$ field follows from 
the effective action as
\begin{gather}
\langle Q^2\rangle=\frac Tm \sum_n\frac{1}{\omega_n^2+\omega_0^2y+2\Gamma|\omega_n|}
\end{gather}
and diverges at $T=0$ as 
$\langle Q^2\rangle\sim |\ln(y)|$. The weak divergence of the fluctuation is a direct consequence of the weakly
 first order nature of the instability at $T=0$.
At finite temperatures, this crosses over to $T/y$ type 
of divergence, as discussed below.

By coupling an external field ($V$) to the position as $VQ$, we can 
study the resulting distortion within linear response.
According to scaling\cite{baxter}, 
\begin{equation}
\langle Q\rangle=V^{1/\delta}\Phi_Q\left(\frac{y}{V^{1/\beta\delta}}\right),
\end{equation}
where $\Phi_Q(x)$ is a scaling function.
From Eq. \eqref{effaction}, we get
\begin{equation}
\langle Q\rangle=\frac{V}{m\omega_0^2y}\sim V^0\left(\frac{y}{V}\right)^{-1},
\end{equation}
from which we deduce $\beta=1/\delta=0$. The scaling function is determined as $\Phi_Q(x)=1/x$. This turns out to be close to 
that 
of the one dimensional Ising model\cite{baxter}.

The divergence of the order-parameter susceptibility defines the $\gamma$ exponent. In the present case, 
the phonon field $Q$ is expected to fluctuate close to the critical coupling $\Gamma_2$, and plays the major role in the instability 
of the system. Its susceptibility is obtained as
\begin{equation}
\chi_Q=\frac{1}{m\omega_0^2}\frac{1}{y}\sim y^{-\gamma}
\end{equation}
with $\gamma=1$. This divergence is analogous to the diverging compressibility at the  WB singularity\cite{varga,loss}.
It gives for the exponent of the external field\cite{cardy} 
\begin{equation}
y_h=\frac{y_t\gamma+d+z}{2}=1.
\end{equation}
Therefore, the universality class of this problem is defined by $d_{eff}=y_t=y_h=1$, as is summarized in Table. 
\ref{exponents}.

\begin{table}[t!]
\centering

\begin{ruledtabular}
\begin{tabular}{c|ccc||ccccc|c}
        & $d_{eff}$ & $y_t$ & $y_h$ & $\alpha$ & $\beta$ & $\gamma$ &  $\delta$ & $\nu$ & $\eta$ \\
\hline
quantum ($T<T^*$) & 1         & 1     & 1     & 1        & 0        & 1   &   $\infty$  & 1 & 1\\
\hline
classical ($T>T^*$) & 0         & 2     & 1     & 2        & -1/2      & 1     &   -1 & - & -
\end{tabular}
\end{ruledtabular}

\caption{Summary of the critical exponent in the quantum and classical critical regions, $\eta$ is only determined through $\eta=d_{eff}+2-2y_h$.}
\label{exponents}
\end{table} 

\section{Relation to the Ising model}

\begin{figure}[h!]
\centering
\centering{\includegraphics[width=3cm,height=3cm]{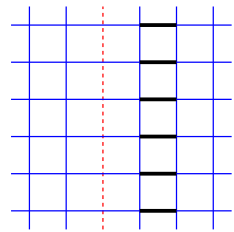}}
\caption{(Color online) Two types of defect chains (red dashed
and thick black lines) are shown for the 2D Ising model, leading to the same local critical behaviour.
\label{2dIsing}}
\end{figure}

Similar exponents characterize the one dimensional classical ferromagnetic Ising model as well. There, for any real $h$ 
(longitudinal magnetic field) and 
finite $T$, there is no sign of criticality. However, at $T=h=0$, the correlation length diverges, indicating the 
presence of a critical point.
Following the ideas of scaling, the exponents were determined\cite{baxter} as $\beta=1/\delta=0$, $\alpha=\gamma=\nu=\eta=1$, 
similarly to what we find here, and the  spatial dimension of this classical model is $d=1$.
Interestingly, this criticality can be brought to finite temperature, if we consider an infinitely strong ferromagnetic 
chain or a line of defects between two neighbouring
columns\cite{bariev,mccoy}
in the two dimensional Ising model on a square lattice, as shown in Fig. \ref{2dIsing}. 
The critical behaviour on the defect chain differs from the bulk critical behaviour, e.g. the critical exponents vary with 
the defect chain strength. In the case of a chain with infinitely strong 
ferromagnetic coupling, the defect contribution to
the specific heat diverges\cite{fisherising} with $\alpha=1$, and the order 
parameter exponent on the defect chain is 
$\beta=0$\cite{mccoy,uzelac}, and the other local critical exponents also agree with that of a one 
dimensional ferromagnetic Ising chain at $T=0$.
Note, that the transition temperature of the two dimensional Ising model is not affected
by the presence of defect line. In this respect, the  effective one-dimensional critical behaviour of the defect is embedded
in a two dimensional critical region, which facilitates its observation.

\section{Correlation functions}

The appearance of quantum criticality is further corroborated by investigating the time 
evolution of the correlator of the phonon 
field, $\langle Q(\tau)Q\rangle$. 
The mapping of our d-spatial dimensional quantum system (with $d=0$) onto a $d+1$ dimensional 
classical one is done in the imaginary time path
integral formalism. There, one introduces an extra imaginary time dimension  with size $1/T$, 
defining the length of the
classical counterpart. Thus, as long as the correlation length, $\xi<1/T$, the system exhibits 
the previous $d+1$
dimensional quantum critical
behaviour. However, for $\xi>1/T$, finite size effects are important, quantum effect can be 
neglected and the system crosses 
over to $d$ dimensional classical problem\cite{suzuki}. 

At $T=0$, the decay of the correlator in imaginary time is obtained from the action, Eq. \eqref{effaction} as 
\begin{gather}
G_Q(\tau)=\langle Q(\tau)Q\rangle\sim \int\limits_{-\infty}^\infty 
\frac{\cos(\omega\tau)\textmd{d}\omega}{2\Gamma_2|\omega|+\omega_0^2y}=\nonumber\\
= \int\limits_0^\infty\frac{\Gamma_2^{-1}\cos(x)\textmd{d}x}{|x|+\omega_0^2y\tau/2\Gamma_2}
\sim\left\{\hspace*{-2mm}\begin{array}{c}
(y\tau)^{-2} \textmd{ for } \tau\gg 2\Gamma_2/\omega_0^2y,  \\
-\ln(y\tau) \textmd{ for } \tau\ll 2\Gamma_2/\omega_0^2y,
\end{array}\right.
\label{gqanal}
\end{gather}
and Eq. \eqref{gqanal} is valid down to $\tau\sim 1/2\Gamma_2$, where the $\omega_{p-}$ 
frequency starts to play its role.
Scaling predicts\cite{cardy} that 
\begin{gather}
G_Q(\tau)=|y|^{2(z-y_h)/y_t}\Phi_\tau\left(\frac{\tau}{y^{-z/y_t}}\right)=\Phi_\tau\left(\tau y\right),
\label{gqscaling}
\end{gather}
where $\Phi_\tau(x)$ is a scaling function, and the last relation is obtained for our specific model. From this, we can draw several 
important conclusions. First, the scaling form predicts a crossover time, separating the long and short $\tau$
regions, to scale as $\tau^* 
\sim 1/y\sim y^{-\nu}$, giving $\nu=1$, 
in agreement with $\nu=1/y_t$. 
This is in perfect agreement with the second integral of Eq. \eqref{gqanal}, where only the $\omega_0^2y\tau/2\Gamma_2$ 
combination 
contains $\tau$ and $y$, suggesting 
\begin{equation}
\tau^*=\frac{2\Gamma_2}{\omega_0^2y}.
\end{equation} By approaching the instability 
($y\rightarrow 0$), 
$\tau^*\rightarrow\infty$.
Second, scaling does not predict additional multiplicative powers of $y$ in front of the scaling function, in nice agreement with the 
analytical result of Eq. \eqref{gqanal}.
Third, the universal scaling function can be determined from Eq. \eqref{gqanal}: $\Phi_\tau(x)$ decreases as $-\ln(x)$ for $x\ll 1$, 
and 
decays 
algebraically as $1/x^2$ for $x\gg 1$. 
The former is characteristic to a quantum Brownian particle\cite{weiss}, which dominates the response when $\tau^*\rightarrow 
\infty$. The latter corresponds to the correlator of a damped harmonic oscillator.
As we approach $y\rightarrow 0$, the harmonic potential flattens, and disappears. Then, our particle does not experience any 
confinement and performs quantum Brownian motion, which is not bounded, and $\langle Q^2\rangle$ diverges. Instead of the 
position 
$Q$, the displacement $Q(\tau)-Q(0)$ keeps track of its dynamics\cite{weiss}.

The electrons also experience criticality similarly to the oscillator. The local charge susceptibility diverges as $1/y$ similarly to
the phononic response\cite{doraphonon}.
The local Green's function of the electrons, $G_e(\tau)=\langle\Psi^+({\bf 0},\tau)\Psi({\bf 0},0)\rangle$ decays as $1/\tau$ for long 
$\tau$ at $T=0$ as in a
local Fermi liquid, but for $\tau<\tau^*$, it changes to
\begin{equation}
G_e(\tau)\sim \frac {1}{2\tau}  \left[1-\left(\frac{\tau}{\tau^*}\right)^2 \exp(2\gamma_E)\right],
\end{equation}
where $\gamma_E\approx 0.577$ is the Euler's constant. Thus, half of the spectral weight is lost at short times due to scattering off 
the phonon close to the critical point $y=0$. Consequently, the spectral function (the Fourier transform of $G_e(\tau)$) takes half 
of its non-interacting value for frequencies larger than $1/\tau^*$. 
This harmonizes with the behaviour of the inelastic scattering rate of 
the electrons\cite{doraelastic}, which reaches its maximal value at $y\rightarrow 0$ for high energy electrons.
The above results are valid for $\tau\gg 1/2\Gamma_2$, below which the other mode 
($\omega_{p-}$) dominates, similarly to Eq. \eqref{gqanal}.

By increasing the temperature, it is better to work in real times since $\tau$ is restricted in a finite slab. 
At $T>0$, $G_Q(t)$ decays exponentially with a correlation time given by $\xi_t=1/2\pi T$. 
This defines the quantum critical 
region, 
when 
besides temperature, there is no other relevant energy scale in the problem. 
The diverging $\xi_t$ at $T\rightarrow 0$ agrees with the algebraic decay of Eq. \eqref{gqanal} at $T=0$.
With increasing temperature, we cross over to the 
classical critical region at 
\begin{equation}
T^*=\frac{\omega_0^2}{4\pi\Gamma_2}y.
\end{equation}
For $T>T^*$, the coherence time becomes $\xi_t=2\Gamma_2/\omega_0^2 y$, as 
we expect by approaching a classical transition.
The very existence of $T^*$ proves the importance of studying quantum criticality. Although we might be away from the critical 
coupling, as long as the relation $T<T^*$ holds, we expect to observe the same quantum critical behaviour as in Eqs. \eqref{gqanal} 
and \eqref{gqscaling}. 
Interestingly,  $T^*$ plays an important role at $T=0$ as well, the two regions of the scaling function are separated by 
$2\pi \tau^*=1/T^*$.
This picture is in perfect accord with critical fluctuation close to $y=0$, which change their nature around $T^*$.

The knowledge of $y_t$ and $y_h$ allows us to formally determine all critical exponents\cite{cardy} 
(see Table \ref{exponents}). These exponents satisfy the scaling as well as quantum 
hyperscaling ($2-\alpha=(d+z)\nu$) relations.

\section{Classical limit}

The high temperature regime of the quantum system can be regarded as the finite sized classical counterpart, with sizes smaller than 
the coherence length.
At finite temperatures, the singular contribution to the free-energy in $y$ can be obtained from Eq. \eqref{freeenQ}, but it is 
instructive to follow a different approach. In the high temperature limit, the harmonic oscillator becomes classical, i.e. $Q$ and 
$P$ being classical variables in Eq. \eqref{hamilton}. Then, after tracing out the electronic degrees of freedom,
we obtain the partition function for the oscillator as
\begin{gather}
Z=\int \textmd{d}P\textmd{d}Q\exp\left(-\frac{P^2}{2mT}-\frac{m\omega_0^2 y Q^2}{2T}-\frac{VQ}{T}\right)=\nonumber\\
=\frac{2\pi T}{\omega_0\sqrt y}\exp\left(\frac{V^2}{2m\omega_0^2 yT}\right),
\end{gather}
where we integrate the exponential of the classical energy over all of phase space (all possible momenta and positions), and we added 
a source term $VQ$.
Then, the singular part of the free-energy is
\begin{equation}
f_s(T\gg T^*)=\frac{T}{2}\ln(y)-\frac{V^2}{2m\omega_0^2y},
\end{equation}
giving $\alpha=2$ and $\gamma=1$. 
This accounts for the instability in the purely classical version of Eq. \eqref{hamilton} for the oscillator, caused by the vanishing 
of the vibration 
frequency. Dissipation has no effect at high $T$, the instability is caused by the potential renormalization. 
At this 
point, we are dealing with a classical harmonic oscillator, therefore $d_{eff}=0$, and the 
critical theory is formally equivalent to that of a "0" dimensional Gaussian model\cite{cardy}. The other exponents are determined as
$\beta=-1/2$ and $\delta=-1$, but the $\nu$ and $\eta$ exponents are senseless, since there is no dimension for the spatial or 
imaginary time dependence. 
We mention that the criticality of the lattice version of Eq. \eqref{hamilton}, the Holstein model is expected to depend on the 
dimensionality of the electrons, and belong to a different universality class.



\section{Summary}

Motivated by the possibility of reaching WB singularity in carbon nanotubes\cite{eggerwb}, we have studied the critical properties 
of a variant of the WB singularity, triggered by local electron-vibration interaction. The 
 only relevant energy scale of the problem is identified as $T^*$, which separates the quantum and classical critical regions. 
The former belongs to the universality class of weakly first order quantum phase transitions, while the latter is formally 
equivalent to a "0" dimensional Gaussian model. 
The two qualitatively different critical regimes are accessible by local vibrational spectroscopy\cite{stipe}, and influence the 
electronic 
response (e.g. dephasing \cite{doraelastic}) as well. 

\begin{acknowledgments}
I'm grateful to M. Gul\'acsi, F. Simon and P. Thalmeier for useful discussions.
This work was supported by the Hungarian
Scientific Research Fund under grant number K72613.
\end{acknowledgments}

\bibliographystyle{apsrev}
\bibliography{wboson}
\end{document}